# Exploiting the Randomness of Large Language Models (LLM) in Text Classification Tasks: Locating Privileged Documents in Legal Matters

Keith Huffman[1]
Chief Legal Counsel
Export Control US (Global)
Head of Export Control Innovation team
SAP
keith.huffman@sap.com

Jianping Zhang
Legal Technology and Data & Analytics
Ankura Consulting Group, LLC
Washington, D.C. USA
jianping.zhang@ankura.com

Nathaniel Huber-Fliflet
Legal Technology and Data & Analytics
Ankura Consulting Group, LLC
London. UK
nathaniel.huber-fliflet@ankura.com

Fusheng Wei
Legal Technology and Data & Analytics
Ankura Consulting Group, LLC
Washington DC, USA
fusheng.wei@ankura.com

Peter Gronvall
Legal Technology and Data & Analytics
Ankura Consulting Group, LLC
Washington DC, USA
peter.gronvall@ankura.com

*Abstract*— In legal matters, text classification models are most often used to filter through large datasets in search of documents that meet certain pre-selected criteria like relevance to a certain subject matter, such as legally privileged communications and attorney-directed documents. In this context, large language models have demonstrated strong performance. This paper presents an empirical study investigating the role of randomness in LLM-based classification for attorney-client privileged document detection, focusing on four key dimensions: (1) the effectiveness of LLMs in identifying legally privileged documents, (2) the influence of randomness control parameters on classification outputs, (3) their impact on overall classification performance, and (4) a methodology for leveraging randomness to enhance accuracy. Experimental results showed that LLMs can identify privileged documents effectively, randomness control parameters have minimal impact on classification performance, and importantly, our developed methodology for leveraging randomness can have a significant impact on improving accuracy. Notably, this methodology that leverages randomness could also enhance a corporation's confidence in an LLM's output when incorporated into its sanctions-compliance processes. As organizations increasingly rely on LLMs to augment compliance workflows, reducing output variability helps build internal and regulatory confidence in LLM-derived sanctions-screening decisions.

*Keywords— AI, eDiscovery, Genai, LLM, Machine Learning, privilege document, randomness and non-determinism of LLM, sanctions compliance,*

I. INTRODUCTION

Text classification is a widely studied area in machine learning, focused on the task of automatically categorizing documents into predefined groups, such as assigning "positive" or "negative" labels in sentiment analysis or in detecting documents as relevant or not relevant to those predefined areas of search, including documents that may be classified as privileged communications between attorneys and clients. This process involves applying a machine learning algorithm to build a predictive model based on a training set of documents with known labels. Once trained, the model can assign categories to new, unreviewed text documents. Text classification is a key topic in Natural Language Processing (NLP) and has numerous real-world applications. Common examples include topic classification and sentiment analysis.

Over the past few years, Large Language Models have progressed at a remarkable pace, transforming from experimental prototypes into powerful, widely adopted tools for a range of language understanding tasks – including text classification. When applying LLMs to text classification, there are two common approaches: In-Context Learning and Fine-Tuning. In-Context Learning uses the model's pre-trained knowledge to make predictions without altering the model's parameters, while Fine-Tuning involves adding new and specifically selected training data to tailor the model to perform a particular task.

LLMs exhibit randomness in text generation, often producing different outputs for the same input. This variability stems from their reliance on probabilistic models, where the model assigns each token (word or sub-word) a probability 'score' reflecting the likelihood that the token will be next given the preceding context. During generation, LLMs use sampling techniques to select the next token based on these probabilities, resulting in diverse and non-deterministic outputs.

Three key parameters – Temperature, Top-P, and Top-K – can be used in combination to control the randomness of LLM outputs.

- Temperature scales the logits (raw output scores) before converting them to probabilities. Higher Temperatures

---

[1]Contributions are author's own

lead to more random and diverse outputs, while lower Temperatures result in more focused and deterministic text.

- Top-P selects tokens from the smallest possible set whose cumulative probability exceeds the threshold specified by Top-P.
- Top-K restricts the selection to the top k most probable tokens, ensuring that only the most likely options are considered during generation.

In the United States, companies responding to litigation or a subpoena from a federal or state enforcement agency (e.g., Department of Justice, Federal Trade Commission, Securities and Exchange Commission, offices of state attorneys general), are obligated to produce to the requesting party all non-privileged material relevant to the legal case. To accomplish this, the company's legal teams are tasked with gathering, compiling, and reviewing large volumes of documents to determine which documents are relevant to the legal case, and then produce copies of those relevant documents in technically-compliant data 'productions.' When attorneys are reviewing the document search of relevant ones, they will also identify which of the documents contain privileged material, so those documents may be withheld from production, under the protections of the attorney-client privilege and its related work-product doctrine. This process is exhaustive and expensive: corporations spend millions or tens of millions of dollars to accomplish this very cumbersome, expensive, and legally-required task. This entire process is referred to in the legal industry as electronic discovery (or 'ediscovery') (Nicholas, et al., 2012) and more specifically, the 'document review' component of e-discovery.

Document review requires significant time and resources to meet production schedules imposed upon corporations by the compulsory legal process. Document review costs continue to escalate as volumes of business data continue to grow. In fact, it can be safely stated that for all practical purposes, data volumes are 'infinite' in size, in the sense that it is essentially impracticable to review each document potentially falling within the scope of an investigation or litigation.

For more than ten years, attorneys have been using machine learning techniques like text classification to more efficiently cull massive volumes of data to identify responsive information. In the legal domain, text classification is typically referred to as predictive coding or technology assisted review (TAR) (Huber-Fliflet, N. et al., 2016).

Our investigation centers on a legal document review task – the central challenge of detecting privileged documents within the ediscovery document review process, so they can be withheld from production. This paper presents an empirical study examining the impact of randomness in LLM-based text classification for privileged document detection during legal document review, with a focus on four key dimensions: (1) performance of LLMs in identifying legal privileged documents, (2) the impact of randomness control parameters on classification outputs, (3) their effect on overall classification performance, and (4) strategies for leveraging randomness to enhance accuracy. Finally, this paper also briefly introduces for further empirical study the impact of randomness in LLM-based support for sanctions compliance.

Section II introduces the privileged document detection task. Section III briefly reviews previous work in text classification using large language models. Section IV outlines our methodology, while Section V presents the experimental results. Finally, Section VI summarizes our findings and outlines our direction for future research

## II. PRIVILEGED DOCUMENT DETECTION

Attorney-client communications and other documents generated at the request of lawyers within legal proceedings are considered 'privileged' documents in today's legal proceedings, and under prevailing procedural rules, are protected from disclosure to requesting parties in litigation and investigation matters. Privileged materials of this nature, as one would expect, are usually emails and electronic documents, either consisting of communications involving lawyers, or documents prepared at the request of lawyers, or otherwise in connection to an actual or anticipated legal matter. Privileged materials are also generally 'protected' from disclosure, by the attorney-client privilege and the work product doctrine, under the procedural rules guiding federal and state legal matters. Because of these important disclosure protections, it is critical for lawyers to screen for these types of documents to ensure that they are removed from document productions provided to requesting parties. The accidental – or 'inadvertent' – production of privileged communications or work product can be potentially devastating to a legal matter. These documents could provide an opposition party with otherwise-unfound insights into a company's proposed legal strategy, regulatory decision-making process, or internal investigation findings.

With data volumes increasing in legal matters, privilege review places a large burden on legal counsel when performing document review – further increasing costs and stressing important deadlines. As a result, legal teams employ numerous workflows specifically designed to undertake privilege document review. Regularly, workflows that target privileged documents employ keyword searching to identify potentially privileged documents allowing attorneys to focus the privilege review population and reduce the volume of documents requiring manual review. Typically, counsel develops a list of potentially privileged keyword terms and applies those terms to the relevant document population to find documents that contain privileged information. These documents are presumed as potentially privileged and are reviewed by counsel to confirm their privilege.

As another example, with global sanctions expanding and data volumes surging, trade compliance teams face mounting pressure to review large sets of both public (e.g., news articles) and non-public documents (e.g., corporate records, payment files) to identify potential sanctions risks. Similar to privilege reviews in litigation, sanctions review workflows increasingly rely on structured methodologies to triage and prioritize high-risk content, as manual review of every document is no longer feasible within tight deadlines.

To manage the rising costs and efforts associated with document review, sanctions teams employ targeted screening techniques, including risk-based filters, predefined keywords (e.g., sanctioned country names, affiliates, or high-risk industries), automated flagging, etcetera to narrow the universe of documents that may require further human review. This empowers compliance teams to prioritize the highest-risk items without compromising accuracy or audit defensibility.

In response to the surging volumes – and incumbent costs – of document review, the legal industry increasingly is turning to machine learning and text classification techniques in search of greater efficiencies and overall accuracy. Most specifically, attorneys and their clients are applying machine learning to build text classification models to identify relevant documents in legal cases. Machine learning techniques have been successful at reducing review populations in legal matters when targeting relevant materials (Huber-Fliflet, N. et al., 2016). These techniques have proved to quantifiably reduce the time and cost of attorney review. While using machine learning to target relevant content has been embraced by the legal community, there unfortunately remains a stigma (or reflexive presumption) that machine learning cannot reliably identify privileged material. Anecdotally, attorneys maintain a belief that a machine learning model is not precise enough to classify privileged material due to the nuance of specific privilege determinations and the relationship-driven nature of communications.

There is little research about the use of machine learning techniques to target privileged information other than (Gabriel, et al. 2013), (Keeling et al, 2019), and (Chhatwal et al, 2020) (Keeling et al., 2022). In these studies, classic text classification techniques are applied to build classification models to identify privileged documents. This paper investigates how leveraging the inherent randomness of LLMs can enhance the performance of privileged document detection.

## III. TEXT CLASSIFICATION WITH LARGE LANGUAGE MODELS

Two common approaches for applying LLMs to text classification tasks are in-context learning and fine-tuning. In-context learning with LLMs (Brown et. al, 2020) involves using prompt engineering to perform new tasks with only a few training or demonstration examples, making it a powerful technique for tasks like text classification (Huber-Fliflet, et al., 2024). In this approach, zero-shot or few-shot learning is often employed.

In zero-shot learning, a model is provided with only a description of the task – without any labeled training examples. Using this description, the model infers how to categorize or classify text. For instance, in sentiment analysis, it may be asked to determine the sentiment of a sentence based solely on the task description.

In few-shot learning, the model is given a handful of labeled examples within the prompt, along with a task description. These examples help the model infer the underlying pattern and apply it to new inputs. For sentiment analysis, the prompt might include several sample sentences paired with their sentiment labels, enabling the model to classify the sentiment of a new target sentence.

In-context learning learns in inference time and does not modify the parameters of an LLM. Even though in-context learning works very well in some applications, it remains unclear how and why in-context learning works. It has become an active research field – researchers are aiming to understand how and why in-context learning works (Xie et al., 2021; Min et al., 2022; Raventós et al., 2023). In-context learning's performance relies heavily on the pre-training corpus as well as the size of the LLM, but more on the pre-training corpus. The pre-training corpus must include text data closely related to the in-context learning task.

Huber-Fliflet et al (2024) presents their work utilizing in-context learning and retrieval-augmented generation (RAG) to identify delay-related statements in the document review process of a construction delay dispute. They found that, specifically, 3-shot learning outperformed zero-shot learning when high-quality, similar examples were available. However, zero-shot learning yielded slightly better results when such examples were lacking. In their work, few-shot learning proved sensitive to the selected examples.

Although LLMs have achieved impressive results across a range of natural language processing tasks, their performance in in-context learning – particularly zero-shot learning – often falls short compared to traditional text classification models trained on huge, labeled datasets. This disparity stems largely from LLMs' limited understanding of category definitions within classification tasks. LLMs tend to perform well only when they fully comprehend the task's nuances. For instance, they often succeed in sentiment analysis because they possess a strong grasp of the task, and the input texts are typically short and straightforward.

Fine-tuning is the process of taking a pre-trained LLM and adapting it to text classification – by training it further on a labeled dataset relevant to that task. Fine-tuning can involve either modifying the parameters of the underlying LLM or adding new classification layers on top of the model and learning task-specific parameters within those layers. By adding new classification layers and fine-tuning, the LLM is transformed into a task-specific classifier.

Fu et al. (2023) conducted a series of experiments to evaluate the impact of fine-tuning DistilBERT, a popular LLM, for text classification. By adding classification layers and fine-tuning the model with domain-specific data, they demonstrated that the adapted DistilBERT consistently outperformed its pretrained counterpart. These findings highlight the critical role of fine-tuning with task-relevant data in enhancing LLM performance for classification tasks.

## IV. PRIVILEGED DOCUMENT DETECTION USING LLMS

The objective of this study is to develop a methodology for leveraging LLMs to identify privileged documents during legal document review. Given the high cost of manual review and the assumption that no human-labeled privileged documents are available, our approach relies on zero-shot learning. LLMs are statistical in nature and incorporate randomness through sampling. As a result, the same document may receive different classifications when submitted to the model multiple times. Our methodology leverages the randomness of LLMs to enhance the correct detection rate of privileged documents.

Our proposed method for leveraging LLM randomness to improve classification performance is straightforward: we submit each document to the model multiple times, generating multiple classification outputs to assess. Due to the inherent variability in LLM responses, a document may be classified as privileged in some instances and not privileged in others. Submitting a privileged document only once risks misclassification; however, repeated submissions increase the likelihood that at least one response correctly identifies it as privileged.

This approach resembles an ensemble method, where each variation in next-token selection effectively simulates a different model. Each time the LLM generates a response with a different token path, it may interpret the document from a distinct perspective or emphasize different textual elements, leading to varied classification outcomes.

By submitting each document multiple times, we assign confidence scores based on classification consistency. A document's confidence score is calculated as the ratio of privileged classifications to the total number of submissions. This ranking enables flexible trade-offs between recall and precision, allowing users to adjust thresholds according to their desired balance.

Multiple submissions can be made using a single randomness control parameter or by varying the parameter across submissions. Higher randomness tends to yield more diverse responses across documents, increasing the likelihood of uncovering alternative classifications. For this reason, we favor higher randomness settings.

For our experiments, we utilized GPT-4.1, which appears to demonstrate a strong grasp of the definition of a privileged document in the context of legal document review. As noted earlier, for LLMs to perform effectively in zero-shot learning scenarios, they must possess a comprehensive understanding of the conceptual definitions underlying the classification tasks. To assess this, we posed the prompt: "*What is a privileged document in legal document review?*" Below is the response provided by GPT-4.1.

> *A privileged document in legal document review refers to any document that is protected from being disclosed to opposing parties due to a recognized legal privilege. The most common types of privilege include:*
>
> 1. **Attorney-Client Privilege**: *This protects communications between a client and their attorney made for the purpose of obtaining legal advice. It ensures that clients can speak freely with their lawyers without fear that those communications will be revealed in legal proceedings.*
> 2. **Work Product Doctrine**: *This protects materials prepared in anticipation of litigation by or for an attorney. It includes notes, memos, and other documents that reflect an attorney's legal theories, strategies, or conclusions.*
> 3. **Confidentiality Privileges**: *These can include other types of communications that are deemed confidential under law, such as doctor-patient or priest-penitent communications.*
>
> *Privileged documents are exempt from discovery, meaning they do not have to be shared with the opposing party, and are protected from being used as evidence in court. Identifying and withholding privileged documents is a critical part of legal document review to protect a client's legal rights and strategies.*

Given GPT-4.1's answer above, we provided it with a straightforward zero-shot prompt to identify privileged documents, followed by the text of the document requiring classification.

> *You are an expert legal analyst. Your task is to evaluate whether the following document is protected by legal privilege. Privileged documents include:*
>
> - *Attorney–Client Communications: Confidential exchanges between a lawyer and their client for the purpose of seeking or giving legal advice.*
> - *Work Product: Materials prepared by or for attorneys in anticipation of litigation.*
>
> *Steps:*
>
> 1. *Read the document below.*
> 2. *Classify it as either "Privileged" or "Not Privileged."*
> 3. *If "Privileged," highlight (quote) one or more passages that demonstrate privilege, noting their locations (e.g., paragraph or line numbers).*
> 4. *Provide a brief rationale (2–3 sentences) explaining your decision*

## V. EXPERIMENTS

We conducted a series of experiments to evaluate our proposed method for privileged document detection using data drawn from six real-world legal document review matters. Specifically, we assessed: (1) the effectiveness of LLMs in identifying privileged documents, (2) their impact on overall classification performance, (3) the influence of randomness control parameters on classification outcomes, and (4) strategies for harnessing randomness to improve accuracy. In this section, we describe the datasets, outline the experimental setup, and present the results of our evaluation.

### A. Dataset

We selected a total of 1,206 documents from six real-world legal document review matters, randomly sampling 201 documents from each. Of these, three documents did not receive responses from the model. All documents contain 1,000 tokens or fewer, with the majority consisting of email communications.

Approximately one-third of the dataset (402 documents) is privileged.

### B. Experimental Setup

All experiments were conducted using GPT-4.1 as the underlying large language model. Classification performance was evaluated using precision and recall. To control the randomness and diversity of generated outputs, three key sampling parameters were considered: Temperature, Top-P, and Top-K. Temperature adjusts the model's confidence by scaling the logits before converting them into probabilities – higher values (up to 2) produce more varied and creative responses, while lower values yield more focused and deterministic text. Top-P (ranging from 0 to 1) enables nucleus sampling by selecting tokens from the smallest set whose cumulative probability exceeds the specified threshold. Top-K limits token selection to the k most probable choices, using a non-zero positive integer. In GPT-4.1, the default values for Temperature, Top-P, and Top-K are 1.0, 1.0, and 50, respectively. In our experiments, we varied the Temperature and Top-P parameters, while maintaining Top-K at its default setting.

The first experiment aimed to evaluate the model's effectiveness in detecting privileged documents. To do this, we submitted the entire document set to GPT-4.1 for classification. The Temperature parameter was set to 0 and Top-P was 0.9, while Top-K was 50. At Temperature = 0, the model is expected to produce the most deterministic and predictable responses.

We conducted two additional experiments to examine how model randomness affects the performance of privileged document detection. In the first experiment, we varied the Temperature parameter across five levels: 0, 0.5, 1, 1.5, and 2, while keeping all other settings constant. We then compared precision and recall metrics at each Temperature setting. The second experiment explored the combined effects of Temperature and Top-P sampling. Temperature values were set to 0, 0.2, 0.5, and 0.9, while Top-P values were set to 0.2, 0.5, and 0.9. We evaluated precision and recall across all combinations of these parameters to assess their joint impact on detection performance.

The goal of the proposed research is to harness the inherent randomness of large language models to enhance the effectiveness of privileged document detection. We aimed to investigate how varying levels of randomness influence classification performance. Higher randomness settings are expected to enhance performance by producing more diverse classification outcomes, thereby increasing the likelihood of identifying more privileged documents and improving recall. In this experiment, each document was submitted to the model five times under each of the previously introduced randomness control settings. Our objective is to confirm that increasing Temperature and Top-P values leads to greater output variability – specifically, a higher rate of documents being classified into different categories across submissions.

In the experiment described above, each document was submitted to the model multiple times, resulting in varying classifications for some documents. For each document, we calculated a confidence score defined as the proportion of submissions classified as privileged. These scores were then used to rank the documents, with higher-ranked entries presumed to have a greater likelihood of being privileged. This ranking approach is expected to enhance the overall effectiveness of privileged document detection by correctly identifying more privileged documents.

### C. Experimental Results

With Temperature set to 0, Top-P at 0.9, and Top-K at 50, the model achieved a precision of 83.4% and a recall of 57.5%. As discussed in Section III, GPT-4.1 demonstrates a strong general understanding of what constitutes a privileged document in the context of legal review. However, its ability to identify only 57.5% of privileged documents suggests limitations in recall, likely due to the absence of matter-specific (or dataset-specific) context. Supporting this, Shani et al. (2025) found that while LLMs effectively form broad conceptual categories aligned with human judgment, they often struggle to capture the nuanced semantic distinctions essential for accurate classification – potentially contributing to the observed recall gap.

For each of the five Temperature settings presented in Table I, every document was submitted to the model five times. The precision and recall values shown are averaged across these submissions. Overall, Temperature variations had minimal impact on classification performance. Notably, the highest precision was observed at a Temperature setting of 2.

TABLE I. PRECISIONS AND RECALLS FOR DIFFERENT TEMPERATURES

| Temperature | Precision | Recall |
|---|---|---|
| 0 | 83.4% | 57.7% |
| 0.5 | 83.0% | 57.6% |
| 1 | 82.9% | 57.7% |
| 1.5 | 83.1% | 57.6% |
| 2 | 83.2% | 58.4% |

Table II presents the average precision and recall values across five submissions for various combinations of Temperature and Top-P settings. Similar to earlier findings, the influence of these randomness control parameters appears minimal. The highest precision and recall were observed at a Temperature of 1 and Top-P of 0.5. Overall, the results from both tables suggest that modest performance gains can be achieved with relatively higher values of Temperature and Top-P.

TABLE II. PRECISIONS AND RECALLS FOR DIFFERENT TEMPERATURES AND TOP-P

| Temperature | Top-P | Precision | Recall |
|---|---|---|---|
| 0 | 0.2 | 83.2% | 57.9% |
| 0 | 0.5 | 83.7% | 58.0% |
| 0 | 0.9 | 83.9% | 58.3% |
| 0.2 | 0.2 | 83.5% | 58.4% |
| 0.2 | 0.5 | 83.4% | 58.3% |
| 0.2 | 0.9 | 83.5% | 58.5% |
| 0.5 | 0.2 | 83.9% | 58.1% |
| 0.5 | 0.5 | 83.8% | 58.7% |
| 0.5 | 0.9 | 83.6% | 58.1% |
| 1 | 0.2 | 83.6% | 58.1% |
| 1 | 0.5 | 84.4% | 59.0% |
| 1 | 0.9 | 83,5% | 58.2% |

We conducted experiments to assess the influence of randomness control parameters – Temperature and Top-P – on the variability of privileged document classification. Specifically, we examined how often a document, when submitted to the model multiple times, would receive differing classifications. As previously noted, each document was submitted five times for classification. Due to the model's inherent randomness, the resulting classifications for a single document could vary, with some classifications labeled as privileged and others not. We hypothesized that Temperature and Top-P would significantly affect this variability: higher values for these parameters were expected to increase the rate of classification inconsistency. In other words, greater randomness settings should lead to more documents receiving divergent classifications across submissions.

Table III and Table IV present the number of documents that received differing classifications across five submissions. We were surprised to find that the randomness control parameters had minimal impact on this variability. Across all settings, approximately 4% of documents were classified inconsistently. In few cases, a document was labeled as privileged for different reasons in separate submissions. These findings align with results reported by Renze and Guven (2024) and Peeperkorn et al. (2024), who similarly observed limited sensitivity to randomness controls in other language processing tasks.

TABLE III. COUNTS OF DOCUMENTS WITH DIFFERENT CLASSIFICATIONS AND DIFFERENT TEMPERATURE SETTINGS

| Temperature | Counts for Docs with different classifications | Counts for Docs with the same classifications |
|---|---|---|
| 0 | 50 | 1,153 |
| 0.5 | 54 | 1,149 |
| 1 | 52 | 1,151 |
| 1.5 | 52 | 1,151 |
| 2 | 54 | 1,149 |

TABLE IV. COUNTS OF DOCUMENTS WITH DIFFERENT CLASSIFICATIONS AND DIFFERENT TEMPERATURE AND TOP-P SETTINGS

| Temperature | Top-P | Counts for Docs with different classifications | Counts for Docs with the same classifications |
|---|---|---|---|
| 0 | 0.2 | 54 | 1,149 |
| 0 | 0.5 | 50 | 1,153 |
| 0 | 0.9 | 50 | 1,153 |
| 0.2 | 0.2 | 50 | 1,153 |
| 0.2 | 0.5 | 47 | 1,156 |
| 0.2 | 0.9 | 51 | 1,152 |
| 0.5 | 0.2 | 48 | 1,155 |
| 0.5 | 0.5 | 52 | 1.151 |
| 0.5 | 0.9 | 47 | 1,156 |
| 1 | 0.2 | 52 | 1,151 |
| 1 | 0.5 | 48 | 1,155 |
| 1 | 0.9 | 47 | 1,156 |

Finally, we present results aimed at improving the effectiveness of privileged document detection by leveraging classification variability across multiple submissions of the same document. For each document, we calculated a confidence score based on the number of times it was classified as privileged. Specifically, for a given document D, the confidence score is defined as:

$$CS(D) = np / ts$$

where np is the number of submissions in which D was classified as privileged, and ts is the total number of submissions. This score reflects the model's consistency in identifying a document as privileged and serves as a basis for ranking documents by their likelihood of being privileged.

Table V and Table VI present precision and recall values across varying cut-off scores for Temperature settings of 0 and 2, respectively. Document scores were calculated using the previously defined formula, based on five submissions per Temperature setting. The results for both configurations are comparable, with Temperature = 2 yielding slightly better performance.

Compared to the precision and recall values from a single submission, five submissions resulted in a 4% increase in recall accompanied by a 1.6% decrease in precision for Temperature = 0, and a 3.5% recall improvement with a 2.4% drop in precision for Temperature = 2. The limited recall gains from five submissions are likely due to the low randomness rate, which remains around 4%. To achieve greater recall improvements, it would be necessary to either increase the number of submissions or raise the randomness rate.

TABLE V. PRECISION AND RECALL VALUES OF 5 SUBMISSIONS WITH TEMPERATURE = 0

| Score Range | Precision | Recall |
|---|---|---|
| 0.90-1.00 | 85% | 53.5% |
| 0.80-0.90 | 84.5% | 55.5% |
| 0.70-0.80 | 84.5% | 55.5% |
| 0.60-0.70 | 83.5% | 57.7% |
| 0.50-0.60 | 83.5% | 57.7% |
| 0.40-0.50 | 82.9% | 60.2% |
| 0.30-0.40 | 82.9% | 60.2% |
| 0.20-0.30 | 81.8% | 61.7% |
| 0.10-0.20 | 81.8% | 61.7% |
| 0.00-0.10 | 33.4% | 100% |

TABLE VI. PRECISION AND RECALL VALUES OF 5 SUBMISSIONS WITH TEMPERATURE = 2

| Score Range | Precision | Recall |
|---|---|---|
| 0.90-1.00 | 86.2% | 54.5% |
| 0.80-0.90 | 84.7% | 56.5% |
| 0.70-0.80 | 84.7% | 56.5% |
| 0.60-0.70 | 83.1% | 58.7% |
| 0.50-0.60 | 83.1% | 58.7% |
| 0.40-0.50 | 81.8% | 60.4% |
| 0.30-0.40 | 81.8% | 60.4% |
| 0.20-0.30 | 80.8% | 61.9% |
| 0.10-0.20 | 80.8% | 61.9% |
| 0.00-0.10 | 33.4% | 100% |

Table VII shows precision and recall values across various cut-off scores based on 12 submissions, each using a distinct combination of Temperature and Top-P settings. This configuration yielded nearly a 6% improvement in recall, with less than a 3% reduction in precision. Table VIII reports results from 25 submissions, derived from five submissions across five different Temperature settings. This setup achieved approximately a 5% gain in recall, accompanied by a 4% drop in precision. Lastly, Table IX summarizes precision and recall for 60 submissions, generated from five submissions across 12 different Temperature and Top-P combinations. This approach resulted in over a 7% improvement in recall, with a modest 3% decrease in precision.

TABLE VII. PRECISION AND RECALL VALUES OF 12 SUBMISSIONS WITH VARYING TEMPERATURE AND TOP-P SETTINGS

| Score Range | Precision | Recall |
|---|---|---|
| 0.90-1.00 | 85.5% | 51.9% |
| 0.80-0.90 | 84.8% | 55.3% |
| 0.70-0.80 | 84.8% | 56.3% |
| 0.60-0.70 | 84.4% | 58% |
| 0.50-0.60 | 83.7% | 59.2% |
| 0.40-0.50 | 83.3% | 59.5% |
| 0.30-0.40 | 82.7% | 60.1% |
| 0.20-0.30 | 82% | 60.4% |
| 0.10-0.20 | 81.6% | 64.1% |
| 0.00-0.10 | 33.4% | 100% |

TABLE VIII. PRECISION AND RECALL VALUES OF 25 SUBMISSIONS WITH VARYING TEMPERATURE AND TOP-P SETTINGS

| Score Range | Precision | Recall |
|---|---|---|
| 0.90-1.00 | 86.2% | 49.8% |
| 0.80-0.90 | 85.1% | 55.5% |
| 0.70-0.80 | 84.3% | 56% |
| 0.60-0.70 | 83.9% | 57% |
| 0.50-0.60 | 83.2% | 58% |
| 0.40-0.50 | 82.7% | 59.5% |
| 0.30-0.40 | 81.8% | 60.4% |
| 0.20-0.30 | 81% | 61.4% |
| 0.10-0.20 | 79.4% | 63.2% |
| 0.00-0.10 | 33.4% | 100% |

The proposed approach not only increases the number of documents correctly identified as privileged but also offers flexibility in balancing recall and precision. For instance, in LLM-based active learning (Xia et al., 2025), documents are initially labeled by a large language model and subsequently used to train a supervised learning model. In this context, precision is prioritized over recall to ensure high-quality training data. As a result, a higher cut-off score may be applied when selecting labeled documents.

TABLE IX. PRECISION AND RECALL VALUES OF 60 SUBMISSIONS WITH VARYING TEMPERATURE AND TOP-P SETTINGS

| Score Range | Precision | Recall |
|---|---|---|
| 0.90-1.00 | 87% | 49.8% |
| 0.80-0.90 | 85.1% | 55.5% |
| 0.70-0.80 | 85.1% | 56.7% |
| 0.60-0.70 | 84.7% | 57.7% |
| 0.50-0.60 | 84% | 58.7% |
| 0.40-0.50 | 83.6% | 59.7% |
| 0.30-0.40 | 82.7% | 60.4% |
| 0.20-0.30 | 82.3% | 61.4% |
| 0.10-0.20 | 81% | 65.7% |
| 0.00-0.10 | 33.4% | 100% |

## VI. CONCLUSIONS

In this paper, we presented an empirical study on the role of randomness in large language models for privileged document detection and introduced a methodology for leveraging this intrinsic variability to enhance text classification performance. Our evaluation focused on four key areas: (1) the effectiveness of LLMs in identifying privileged documents, (2) their impact on overall classification metrics, (3) the influence of randomness control parameters on classification consistency, and (4) strategies for harnessing randomness to improve precision and recall.

The experimental results revealed that the large language model was able to identify fewer than 60% of privileged documents, achieving a precision of approximately 84%. We found that randomness control parameters had minimal influence on classification performance and did not significantly alter the underlying randomness rates.

However, our proposed methodology demonstrated its potential to improve recall, albeit with a slight reduction in precision. The extent of recall enhancement was constrained by the low inherent randomness rates. Notably, increasing the number of submissions led to more substantial gains in recall. Given how sensitive privileged information is, these recall gains, with minimal impact on precision, are impactful and important. They add rigor to the process, increase accuracy, reduce legal risk, and reduce costs: each of these elements are foundational to meeting the legal objectives in litigation and investigation matters. We conclude and propose that this technique can be used to help improve the process of identifying privileged documents with LLMs.

In the future, we plan to conduct additional experiments on larger datasets. We will continue to explore methodologies for harnessing the inherent randomness of LLMs to enhance the effectiveness of text classification. In particular, we will investigate techniques for increasing randomness rates and evaluate the performance of various LLM architectures. Beyond leveraging randomness, we will also explore ensemble approaches – combining classification outputs from different prompts and multiple models – to further improve overall classification accuracy and robustness. The experiments utilized documents from six distinct legal matters. The authors suggest that conducting experiments on a matter-by-matter basis and incorporating matter-specific knowledge into the prompts could yield significantly improved results.

## VII. FUTURE WORK ON SANCTIONS COMPLIANCE

Sanction compliance ensures that a business or institution adheres to economic and trade sanctions imposed by governments or international bodies. Sanction compliance is critically important for organizations operating in global markets. This section outlines future work exploring how the inherent randomness of large language models can be leveraged to enhance sanction compliance.

### A. Sanctions Compliance Overview

Economic sanctions have become increasingly employed tools for advancing foreign policy, national security and other objectives. Intergovernmental bodies such as the United Nations, as well as individual governments, frequently impose sanctions on designated companies, organizations, and individuals. These measures typically restrict certain commercial or financial activities, either comprehensively or in narrowly defined contexts. In some instances, governments will even impose so-called counter-sanctions, which are measures imposed by a government to sanction individuals, entities, or sectors in response to that foreign government's sanctions, effectively retaliating by restricting their activities.

To comply with these regimes, companies are expected to conduct sanctions screening to determine whether their business partners, customers, or counterparties are a designated party. Sanctions authorities publish lists of these designated parties, often accompanied by identifying details such as name, address, nationality, and other attributes. However, because many sanctioned individuals and companies share common or identical names with non-sanctioned parties, companies usually evaluate additional identifiers, such as date of birth, country, or registered address, to distinguish true matches from false positives. A name match alone signals only a potential link, not a confirmed identity, and typically requires additional data to resolve accurately.

Although this matching process can be performed manually, most organizations employ automated screening systems. Sanctions lists are lengthy, frequently updated, and must be screened against large volumes of data, making manual review both inefficient and error-prone. Automation ensures consistent, real-time screening, supports comprehensive record-keeping, and reduces the likelihood of missing a valid match, an especially important consideration given the severity of potential consequences. Sanctions violations can result in substantial fines, exclusion from markets, loss of licensing privileges, reputational harm, and, in some jurisdictions, criminal liability.

### B. The Business Problem

Traditional screening tools, however, often rely on basic name-matching logic and therefore generate high rates of false positives. In response, companies are slowly but increasingly adopting machine learning and generative AI methodologies to improve efficiency and accuracy. Machine learning and large language models can evaluate and extract contextual information, such as addresses, corporate affiliations, language, or transactional behavior to more effectively differentiate between sanctioned and non-sanctioned parties. Likewise, machine learning and large language models can analyze unstructured data sources, including media reporting or corporate disclosures, to uncover hidden risk indicators and help resolve potential sanctioned party list matches. These innovations reduce manual review burdens, accelerate decision-making, and strengthen the overall effectiveness of sanctions compliance programs.

Despite clear benefits, Some compliance teams remain cautious about widespread adoption of machine learning due to concerns over the unpredictability of its outputs. One illustrative scenario involves sanctions screening: a model may correctly identify that an entity appears on a "sanctions list," but fail to distinguish that the designation was made by a counter-sanctions jurisdiction rather than a jurisdiction relevant to the organization. For example, consider an LLM asked whether Company A is sanctioned. The company is legally established in Country Y but was designated by Country X in retaliation to sanctions actions by Country Y. While that measure by Country X may not have legal or operational consequences for companies operating in Country Y, a model returning a "true match" without contextual nuance could lead a company operating in Country Y to incorrectly block or refuse business with Company A. This highlights how misclassification, especially in the context of increasingly conflicting geopolitics, can generate unnecessary business disruption and compliance challenges.

### C. Future Sanctions Screening Experiments

To address the problem of unpredictable output from large language models in sanctions screening, particularly the challenge of distinguishing relevant sanctions from counter-sanctions, we would explore an experiment that improves reliability, transparency, and confidence in LLM-based solutions. The core objective is to ensure that when an LLM identifies a company as "sanctioned," it can also correctly determine which jurisdiction's sanctions matter for the corporation. This matters because a model may correctly identify that a company appears on a sanctions list, but without proper context, could identify a company solely sanctioned by a counter-sanctions regime as a compliance risk for a company not subject to the counter-sanctions measures, for example.

The experiment would combine retrieval-augmented generation with structured reasoning and controlled randomness to strengthen accuracy. First, the model would retrieve official regulatory information from specified jurisdictions and extract explicit "evidence spans" showing where a designation appears. That evidence would then be passed to a large language model that applies a defined policy (e.g., "treat only XYZ sanctions as relevant; counter-sanctions are informational only") and assign a classification label. To increase confidence and reduce natural randomness, this process could be repeated multiple times with low-variance sampling and a vote distribution is generated, enabling the system to report a calibrated confidence score rather than a binary yes/no answer.

This approach could not only strengthen the trustworthiness of automated sanctions screening but also give compliance teams an audit trail: the model would cite specific regulatory text, clearly identify the jurisdiction, abstain when uncertain, and quantify confidence in its conclusion. Future research can expand this work beyond sanctions to other high-stakes compliance domains, such as export controls or anti-corruption risk, where geopolitical divergence and unique compliance requirements increasingly require explainable, evidence-backed automation rather than vague classification outputs.